\documentclass[12pt, leqno]{article}

\usepackage[letterpaper,top=2cm,bottom=2cm,left=3cm,right=3cm, marginparwidth=2cm]{geometry}

\usepackage{amsthm}
\usepackage{graphicx}
\usepackage{microtype}
\usepackage[usenames,dvipsnames]{xcolor}
\usepackage{amsmath,amssymb,amsfonts,latexsym,cancel}
\usepackage{amsmath, amstext, amsfonts}
\usepackage{float}
\usepackage{caption}
\usepackage{appendix}
\usepackage{multirow}

\usepackage{cite}

\usepackage{subfig}
\usepackage{tabularx}
\usepackage{algorithm}
\usepackage[noend]{algpseudocode}

 \usepackage{mathptmx}      

\newtheorem{theorem}{Theorem}[section]

\begin{document}

{\centering \LARGE An image encryption algorithm based on chaotic Lorenz system and novel primitive polynomial S-boxes  \par}
\bigskip

{\centering \normalsize Temadher Alassiry Al-Maadeed$^1$, Iqtadar Hussain$^1$, Amir Anees$^2$, M. T. Mustafa$^1$ \par}
{\centering\itshape $^{1}$Department of Mathematics, Statistics and Physics, Qatar University, Doha, 2713, Qatar \par}
{\centering\itshape $^{2}$Department of Computer Science and Information Technology, La Trobe University, Melbourne, Australia \par}
{\centering\itshape  t.alassiry@qu.edu.qa,  iqtadarqau@qu.edu.qa, a.anees@latrobe.edu.au, tahir.mustafa@qu.edu.qa   \par}

\begin{abstract}
Nowadays, the chaotic cryptosystems are gaining more attention due to their efficiency, the assurance of robustness and high sensitivity corresponding to initial conditions. In literature, on one hand there are many encryption algorithms that only guarantee security while on the other hand there are schemes based on chaotic systems that only promise the uncertainty. Due to these limitations, each of these approaches cannot adequately encounter the challenges of current scenario. Here we take a unified approach and propose an image encryption algorithm based on Lorenz chaotic system and primitive irreducible polynomial S-boxes. First, we propose 16 different S-boxes based on projective general linear group and 16 primitive irreducible polynomials of Galois field of order 256, and then utilize these S-boxes with combination of chaotic map in image encryption scheme. Three chaotic sequences can be produced by the Lorenz chaotic system corresponding to variables $x$, $y$ and $z$. We construct a new pseudo random chaotic sequence $k_i$ based on $x$, $y$ and $z$. The plain image is encrypted by the use of chaotic sequence $k_i$ and XOR operation to get a ciphered image. To demonstrate the strength of presented image encryption, some renowned analyses as well as MATLAB simulations are performed. \\

\noindent
{\bf Keywords: }   Lorenz System, Chaos,  Substitution box, image encryption, cryptanalysis.
\end{abstract}

\section{Introduction}
\label{intro}
Last decade is considered as a remarkable era for secure communication and image processing. In wireless communication, protection of digital data like text, sound, image and video has increased importance because multimedia elements have taken hold on many important fields like electronic commerce, banking industry, law enforcement agencies requirements and personal data. The performance of old cryptosystems for image encryption is poor in encryption of bulk sized data \cite{r1, r2}. To tackle this problem, new schemes based on chaos for image encryption have been developed. In \cite{r3, r4}, Amigo et. al., and Jakimoski, have shown a link between secure communication and chaos theory. They said chaotic maps could achieve some basic requirements of secure communication like randomness, robustness and sensitivity to initial conditions. In \cite{r5, r6}, Ott and Alvarez have observed that values breaded by chaotic maps can be regained based on initial conditions but extremely erratic, and this kind of behavior is valuable for cryptosystems. Based on these properties, some cryptographers have proposed novel cryptosystems in \cite{r7, r8}. Pseudorandom number generator based on chaotic maps is one of the emerging field nowadays and can be utilized in different cryptosystems to get more security \cite{r9, r10}.

Substitution box (S-box) is a fundamental component in symmetric key algorithms which implement substitution. Substitution boxes are building blocks of symmetric cryptosystems. The substitution tables (S-boxes) play a crucial part in the encryption algorithms in order to meet the definition of a perfect security. Sometime big structures like S-box slowdown the processing of the encryption in a scenario where big data processing is required, the main reason is the complexity of that system and this kind of negative effect reduce the utility of that encryption algorithm in practical communication. In literature, there are many chaotic schemes that are using one or multiple (S-boxes) to get more security. There is no doubt that when one will use S-box, the confusion creating ability of that algorithm between ciphertext and secret key will improve but it will definitely reduce the encryption speed \cite{r11, r12, r13, r14, r15, r16, r17, r18, r19, r20, r21, r22, r23, r24, r25, r26, r27, r28, r29}. Here, we propose an image encryption algorithm that builds on chaotic maps and S-boxes. First, we propose novel 16 different S-boxes based on 16 different primitive irreducible polynomials of Galois field of order 256 and projective general linear group, then the proposed S-boxes are utilized in image encryption. The focus is to construct an efficient and secure image encryption scheme building on straightforward and less complex steps, employing chaotic Lorenz system.

Some basic definitions of mathematical background for cryptography and details of chaotic map are given in Section 2. The detailed description of  image encryption algorithm proposed in this work is given in Section 3. Section 4 depicts the outcomes of simulation and different analyses. The conclusion of whole scheme is given in Section 5. 

\section{Basic Definitions}
The structural units of designed image encryption technique are briefly discussed in this section. First, an introduction to the Galois field and its primitive irreducible polynomials is presented which is followed by the basics of Projective General Linear Group. Next, the comprehensive description of chaotic Lorenz map is given.

\subsection{Galois field and its primitive irreducible polynomials}
It is known from the theory of Galois field that  if $p$ is a non-zero element of a Principle Ideal Domain (PID) $\textbf{R}$, then  $\frac{R}{p}$ will be a field if $p$ is irreducible. Therefore, for a prime $p$ and $q=p^n$, the finite field of order $q$ can be given as 
$\textbf{GF}(q)=\textbf{GF}(p^n)$. 
The polynomial extension $\textbf{R}[x]$ of intergral domanin $\textbf{R}$ is also intergral domain, therefore, in case of polynomial extension $\textbf{R}[x]/\left\langle p(x)\right\rangle$ will be a field structure when $p(x)$ is primitive irreducible polynomial, where $\left\langle p(x)\right\rangle$ is a maximal ideal.
\begin{eqnarray}
 \textbf{GF}(q)=\frac{\textbf{GF}(p)[x]}{\left\langle m(x)\right\rangle}
\end{eqnarray}
with $m(x)$ denoting a monic primitive irreducible polynomial of degree $n$ in Galois field $\textbf{GF}(p^n)$. An example of above formula is as follows:
 \begin{eqnarray}
 \textbf{GF}(2^8)=\frac{\textbf{GF}(2)[x]}{\left\langle x^8+x^7+x^6+x^5+x^4+x^2+1 \right\rangle}
\end{eqnarray}
\begin{eqnarray*}
 \textbf{GF}(2^8)=\left\{a_1+a_2x+a_3x^2+a_4x^3+a_5x^4+a_6x^5+a_7x^6+a_8x^7+\left\langle 
m(x)\right\rangle|a_i\in \textbf{GF}(2)\right\}
\end{eqnarray*}
The elements of $\textbf{GF}(2^8)$ can be represented by a polynomial of degree 8 and $\left\langle m(x)\right\rangle$ is the maximal ideal generated by monic irreducible polynomial, when the degree of polynomial will exceed from 7 this maximal ideal will absorb it. Now, the question is how many different irreducible polynomials are there corresponding to any Galosi field $\textbf{GF}(p^n)$. The formula to find all irreducible polynomials is as follows:
\begin{eqnarray}\label{eq4}
 \frac{1}{n} \stackrel{d/n}{\sum} \mu(d) P^{n/d}
\end{eqnarray}
By using above formula, it can be seen in Table 1, that there are 30 irreducible polynomials for $\textbf{GF}(2^8)$. But for the construction of Galosi field which can generate its non-zero elements we need primitive irreducible polynomials. In Table 1, we have shown that out of 30 irreducible polynomials 16 are primitive irreducible. We have used Rabin's test to find 30 irreducible polynomials, Rabin's test is as follows:
\begin{theorem} \label{thm1}
Let $p_1,p_2,...,p_k$ be all the prime divisors of $n$, and denoted $n_i=n/p_i$, for $1 \leq i\leq k$. A polynomial $f \in \textbf{F}_q[x]$ of degree $n$ is irreducible in $\textbf{F}_q[x] \Leftrightarrow gcd(f, x^{q^{n_i}}-x mod f)=1$ for $1 \leq i\leq k$, and $f$ divides $x^{q^n}-x$.
\end{theorem}

A  polynomial that generates all elements of an extension field from a base field is called a primitive polynomial. It is worth noting that the primitive polynomials are irreducible polynomials as well. There exists a primitive polynomial of degree $n$ over $\textbf{GF(q)}$ for any prime or prime power $q$ and any positive integer $n$. There are
\begin{eqnarray}
 a_q(n)=\frac{\phi(q^n-1)}{n}
\end{eqnarray}
primitive polynomials over $\textbf{GF}(q)$, where $\phi(n)$ is the totient function.
\begin{theorem}
A polynomial of degree $n$ over the finite field $\textbf{GF}(2)$ is primitive if it has polynomial order $2^n-1$.
\end{theorem}

\begin{table*}
\centering 
\begin{tabular}{l | c | c } 
\hline\hline
 8 degree polynomials of $\textbf{GF}(2^8)$ & Irreducible & Prim. irreducible \\ 
\hline

$x^8+x^7+x^5+x^4+1$ & Yes & No\\

$x^8+x^4+x^3+x^2+1$ & Yes & Yes\\

$x^8+x^5+x^3+x^1+1$ & Yes & Yes\\
$x^8+x^7+x^6+x^4+ x^3+ x^2+1$ & Yes & No\\

$x^8+x^6+x^5+x^4+ x^2+ x^1+1$ & Yes & No\\

$x^8+x^7+x^6+x^5+ x^4+ x^3+1$ & Yes & No\\

$x^8+x^5+x^3+x^2+1$ & Yes & Yes\\

$x^8+x^6+x^4+x^3+x^2+x^1+1$ & Yes & Yes\\

$x^8+x^4+x^3+x+ 1$ & Yes & No\\

$x^8+x^7+x^6+x^1+1$ & Yes & Yes\\

$x^8+x^6+x^5+x^2+1$ & Yes & Yes\\

$x^8+x^6+x^5+x^4+x^3+x^1+1$ & Yes & No\\

$x^8+x^7+x^2+x^1+1$ & Yes & Yes\\

$x^8+x^7+x^5+x^4+x^3+x^2+1$ & Yes & No\\

$x^8+x^7+x^3+x^2+1$ & Yes & Yes\\

$x^8+x^7+x^6+x^5+x^4+x^2+1$ & Yes & Yes \\

$x^8+x^5+x^4+x^3+ x^2+ x^1+1$ & Yes & No\\

$x^8+x^7+x^6+x^5+x^2+x^1+1$ & Yes & Yes\\
$x^8+x^7+x^6+x^4+ x^2+ x^1+1$ & Yes & No \\

$x^8+x^6+x^3+x^2+1$ & Yes & Yes\\

$x^8+x^7+x^4+x^3+ x^2+ x^1+1$ & Yes & No\\

$x^8+x^7+x^6+x^3+x^2+x^1+1$ & Yes & Yes\\

$x^8+x^7+x^6+x^5+ x^4+ x^1+1$ & Yes & No\\

$x^8+x^6+x^5+x^1+1$ & Yes & Yes\\

$x^8+x^5+x^4+x^3+1$ & Yes & No\\

$x^8+x^6+x^5+x^3+1$ & Yes & Yes\\

$x^8+x^7+x^5+x^1+1$ & Yes & No\\

$x^8+x^6+x^5+x^4+1$ & Yes & Yes\\

$x^8+x^7+x^3+x^1+1$ & Yes & No\\

$x^8+x^7+x^5+x^3+1$ & Yes & Yes\\

\hline

\hline
\end{tabular}
\caption{Irreducible and primitive irreducible polynomials corresponding to $\textbf{GF}(2^8)$.} 
\label{t_c} 
\end{table*}

Theorem \ref{thm1},yields 30 irreducible polynomial of Table 1. The next question is how to get 16 primitive irreducible polynomials from these 30 irreducible polynomials. The procedure is explained through an example of $GF(2^4)$.
\begin{eqnarray}
 \textbf{GF}(2^4)=\frac{\textbf{GF}(2)[x]}{\left\langle x^4+x^3+1\right\rangle}
\end{eqnarray}
There are two primitive irreducible polynomials for $\textbf{GF}(2^4)$. The process to check whether an irreducible polynomial is primitive irreducible or not is shown in the example and counter example below.

 Suppose $f(x)= x^4+x^3+1$ is an irreducible polynomial. Let $\alpha$ be the root of $f(x)$. If $\alpha$ is the root of $f(x)$ then we have
\begin{eqnarray}
f(\alpha)={\alpha}^4+{\alpha}^3+1=0
\end{eqnarray}
\begin{eqnarray}\label{eq8}
{\alpha}^4={\alpha}^3+1
\end{eqnarray}
Because the coefficients of polynomial are in $\textbf{GF}(2)$, that is why $-1=+1$.
\begin{eqnarray}\label{eq9}
{\alpha}^5={\alpha}^4+\alpha={\alpha}^3+1+\alpha={\alpha}^3+\alpha+1
\end{eqnarray}
\begin{eqnarray}\label{eq10}
{\alpha}^6={\alpha}^4+{\alpha}^2+\alpha={\alpha}^3+1+{\alpha}^2+\alpha={\alpha}^3+{\alpha}^2+\alpha+1
\end{eqnarray}
\begin{eqnarray}\label{eq11}
{\alpha}^7={\alpha}^4+{\alpha}^3+{\alpha}^2+\alpha={\alpha}^3+1+{\alpha}^3+{\alpha}^2+\alpha={\alpha}^2+\alpha+1
\end{eqnarray}
Where $2{\alpha}^3=0$ due to $GF(2)$.
\begin{eqnarray}
{\alpha}^8={\alpha}^3+{\alpha}^2+\alpha
\end{eqnarray}
\begin{eqnarray}
{\alpha}^9={\alpha}^4+{\alpha}^3+{\alpha}^2={\alpha}^3+1+{\alpha}^3+{\alpha}^2={\alpha}^2+1
\end{eqnarray}
\begin{eqnarray}
{\alpha}^{10}={\alpha}^3+\alpha
\end{eqnarray}
\begin{eqnarray}
{\alpha}^{11}={\alpha}^4+{\alpha}^2={\alpha}^3+1+{\alpha}^2={\alpha}^3+{\alpha}^2+1
\end{eqnarray}
\begin{eqnarray}
{\alpha}^{12}={\alpha}^4+{\alpha}^3+\alpha={\alpha}^3+1+{\alpha}^3+\alpha=\alpha+1
\end{eqnarray}
\begin{eqnarray}
{\alpha}^{13}={\alpha}^2+\alpha
\end{eqnarray}
\begin{eqnarray}
{\alpha}^{14}={\alpha}^3+{\alpha}^2
\end{eqnarray}
\begin{eqnarray}
{\alpha}^{15}={\alpha}^4+{\alpha}^3={\alpha}^3+1+{\alpha}^3=1
\end{eqnarray}
It can be seen in above example that we are getting ${\alpha}^{15}=1$, and the order of $\textbf{GF}(2^4)$ is $16$.  This means $f(x)= x^4+x^3+1$ is a primitive polynomial because it is generating all non-zero elements of $\textbf{GF}(2^4)$. Here $\alpha$, the root of primitive polynomial, is known as primitive element.   In other words, because $\textbf{GF}$ is also a cyclic group so $\alpha$ is the generator. All irreducible polynomials are not primitive, to show this fact a counter example is given below.

Suppose $f'(x)= x^4+x^2+1$ is an irreducible polynomial. Let $\beta$ be the root of $f(x)$. If $\beta$ is the root of $f'(x)$ then we have
\begin{eqnarray}
f'(\beta)={\beta}^4+{\beta}^2+1=0
\end{eqnarray}
\begin{eqnarray}
{\beta}^4={\beta}^2+1
\end{eqnarray}
Because the coefficients of polynomial are in $\textbf{GF}(2)$, that is why $-1=+1$.
\begin{eqnarray}
{\beta}^5={\beta}^3+\beta
\end{eqnarray}
\begin{eqnarray}
{\beta}^6={\beta}^4+{\beta}^2={\beta}^2+1+{\beta}^2=2{\beta}^2+1=1
\end{eqnarray}
It can be seen that $f'(x)= x^4+x^2+1$ is irreducible but not primitive, because it is not generating all non-zero elements of $\textbf{GF}(2^4)$. Similarly, in Table 1, we obtain all primitive irreducible polynomials form irreducible polynomials. 

\subsection{Projective General Linear Group (PGL)}
The Projective General Linear Group (PGL) defined as the group acting on $\stackrel{-}{F}=\textbf{GF}(p^n)\bigcup \left\{\infty\right\}$ is the group of all transformations and is denoted by $PGL(2,\textbf{GF}(p^n))$ . With the standard understanding about $\infty$, $PGL(2,\textbf{GF}(p^n))$  is the set of all linear fractional transformations (LFT) of $\stackrel{-}{F}=\textbf{GF}(p^n)\bigcup \left\{\infty\right\}$, 
\begin{equation}
PGL(2,\textbf{GF}(p^n)) =\left\{g: \stackrel{-}{F}\longrightarrow \stackrel{-}{F} |  g(z)=\frac{az+b}{cz+d}, {a,b,c,d}\in \textbf{GF}(p^n), \  ad-bc\neq 0 \right\}
\end{equation}
Linear fractional transformation $g(z)$ is shortly denoted by LFT. We study a special class of maps
\begin{eqnarray}
f: PGL(2,\textbf{GF}(2^8))\times \textbf{GF}(2^8)\longrightarrow \textbf{GF}(2^8)
\end{eqnarray}
A LFT of $PGL(2,\textbf{GF}(2^8))\times \textbf{GF}(2^8)$ is a map of the form $g(z)=\frac{az+b}{cz+d},$ ${a,b,c,d}\in \textbf{GF}(2^8)$ with $ad-bc\neq 0$

This transformation is depending on the invertible $2\times 2$ matrix 
$\begin{pmatrix} 
a & b\\ 
c & d
\end {pmatrix}$

\subsection{Chaotic Lorenz System}
The idea of defining the chaotic dynamics with the help of chaotic maps is a big breakthrough in the field of dynamical systems. The mathematical modelling related to atmospheric convection was presented by E. Lorenz \cite{r13, r14}  through the system of chaotic differential equation given by
\begin{eqnarray}
  \frac{dx}{dt} &=& a(y-x) \label{eq:X}\\
  \frac{dy}{dt} &=& bx - y -xz \label{eq:Y}\\
  \frac{dz}{dt} &=& xy - cz \label{eq:Z}
\end{eqnarray}
with the variables $x$,$y$ and $z$ in the intervals $-60 \leq x \leq 60$, $-60 \leq y \leq 60$, $-60 \leq z \leq 60$. For chaotic behavior, the values for parameters $a$,$b$, and $c$ respectively are $a=10$, $b=28$ and $c=8/3$.

\section{Primitive irreducible polynomial S-boxes}
	Now from the above linear transforamtion, we have;
	\begin{eqnarray*}
f_i: PGL\left(2,GF(2^8)=\frac{GF(2)[x]}{\left\langle p_i(x) \right\rangle}\right)\times \left(GF(2^8)=\frac{GF(2)[x]}{\left\langle p_i(x)  \right\rangle}\right)\longrightarrow \left(GF(2^8)=\frac{GF(2)[x]}{\left\langle p_i(x)  \right\rangle}\right)
\end{eqnarray*}
where $p_i(x)$, $i=1,2,3,...,16$ are set of primitive irreducible polynomials of Table 1 for $GF(2^8)$. Therefore, we have 16 different $f_i$, where $i=1,2,3,...16$ to construct 16 different S-boxes with fixed $a,b,c,d \in GF(2^8)$. The order of $PGL\left(2,GF(2^8)\right)$ is $16776960$, therefore one can construct huge number of S-boxes by changing $a,b,c,d \in GF(2^8)$. In this section, we have given an example of creating one S-box using $a=32,b=22,c=11,d=8 \in GF(2^8)$ and $p_1(x)=x^8+x^4+x^3+x^2+1$. In polynomial form, $a=x^5, b=x^4+x^2+x, c=x^3+x+1, d=x^3$. Here it must be noted that the  sign $‘+’$ indicates XOR operation.
\begin{eqnarray}
f_1(z)=\frac{(x^5)(z)+(x^4+x^2+x)}{(x^3+x+1)(z)+(x^3)} . 
\end{eqnarray}
For $z=0$
\begin{eqnarray*}
f_1(0)=\frac{(x^5)(0)+(x^4+x^2+x)}{(x^3+x+1)(0)+(x^3)}=\frac{x^4+x^2+x}{x^3}=\frac{\mu^{239}}{\mu^3}=\mu^{239-3-1}=\mu^{237}=237
\end{eqnarray*}
where $\mu^{239}=\mu^4+\mu^2+\mu$, $\mu^3=x^3$ based on $p_1(x)=x^8+x^4+x^3+x^2+1$. It is to be noted that corresponding to different primitive polynomials $p_i(x)$ of $GF(2^8)$ these values of $\mu$ power will be different.

For $z=1$
\begin{eqnarray*}
f_1(1)=\frac{(x^5)(1)+(x^4+x^2+x)}{(x^3+x+1)(1)+(x^3)}=\frac{x^5+x^4+x^2+x}{2x^3+x+1}=\frac{x^5+x^4+x^2+x}{x+1}=\frac{\mu^{249}}{\mu^{25}}=\mu^{249-25-1}=\mu^{225}=225
\end{eqnarray*}
where $\mu^{249}=\mu^5+\mu^4+\mu^2+\mu$, $\mu^{25}=\mu+1$ based on $p_1(x)=x^8+x^4+x^3+x^2+1$

For $z=2$
\[ f_1(2)=\frac{(x^5)(2)+(x^4+x^2+x)}{(x^3+x+1)(2)+(x^3)}=\frac{(x^5)(x)+(x^4+x^2+x)}{(x^3+x+1)(x)+(x^3)}=\frac{x^6+x^4+x^2+x}{x^4+x^3+x^2+x} \]

\begin{eqnarray}
=\frac{x^5+x^4+x^2+x}{x+1}=\frac{\mu^{219}}{\mu^{76}}=\mu^{219-76-1}=\mu^{144}=144
\end{eqnarray}

where $\mu^{219}=\mu^6+\mu^4+\mu^2+\mu$, $\mu^{76}=\mu^4+\mu^3+\mu^2+\mu$ based on $p_1(x)=x^8+x^4+x^3+x^2+1$. Similarly, for $z=3,4,5,...,255$ all the elements can be constructed corresponding to primitive irreducible polynomial $p_1(x)=x^8+x^4+x^3+x^2+1$.  
Table 2 displays all the elements of  $p_1(x)$ S-box. 

\begin{table}[h] 
\centering 
\resizebox*{.95\textwidth}{!}{
\begin{tabular}{c | c | c | c | c | c | c | c | c | c | c | c | c | c | c | c | c} 
\hline\hline
 & 0 & 1 & 2 & 3 & 4 & 5 & 6 & 7 & 8 & 9 & 10 & 11 & 12 & 13 & 14 & 15  \\ 
\hline 
0&237&225&144&236&211&25&147&20&185&127&132&195&123&136&197&170\\
\hline
1&109&112&61&84&183&4&186&54&234&121&177&129&215&48&41&1\\
\hline
2&162&228&194&150&141&175&74&91&70&50&47&85&176&40&34&102\\
\hline
3&119&223&202&206&7&22&98&158&190&148&69&30&38&113&179&224\\
\hline
4&131&104&165&178&106&169&174&116&26&154&21&90&65&157&76&64\\
\hline
5&45&5&253&86&172&124&180&67&247&115&42&118&217&240&189&192\\
\hline
6&199&12&6&125&216&254&251&231&210&227&126&160&151&107&73&139\\
\hline
7&77&122&188&8&16&232&153&111&143&203&24&39&95&99&78&182\\
\hline
8&89&213&241&171&81&9&72&13&105&205&3&59&120&245&35&168\\
\hline
9&137&27&66&97&79&71&55&226&201&187&214&239&80&2&208&255\\
\hline
10&63&156&249&135&83&248&110&140&29&163&155&219&184&49&68&173\\
\hline
11&200&10&149&51&23&57&157&14&94&58&15&209&18&103&193&142\\
\hline
12&133&11&56&181&242&43&96&196&33&229&37&220&130&60&88&212\\
\hline
13&46&93&44&221&62&87&114&100&75&246&230&222&204&235&19&164\\
\hline
14&128&233&252&117&82&146&138&17&161&191&53&218&166&52&145&23\\
\hline
15&159&108&198&28&92&31&243&207&32&134&244&0&250&152&36&101\\
\hline 
\end{tabular}
}
\caption{ The Proposed S-box} 
\label{t_b} 
\end{table}

\subsection{Analysis for assessing the  S-box strength}
The analysis section focuses on determining the cryptographic strength of the designed S-box through appropriate standard checks. 
The strength analysis of the designed S-boxes is carried out by employing the commonly used standard criteria constituting bijectivity, nonlinearity, bit independence criterion (BIC), strict avalanche criterion (SAC), differential approximation probability (DP), and  linear approximation probability (LP). The reader is referred to 
\cite{ranalysis} for definitions and important characteristics of these criteria. Table 3 presents the outcome of the strength analsyes for our proposed S-boxes with different Primitive polynomial of $GF(2^8)$.

%
%
%
%
%
\begin{table*}[h]
\centering 
\resizebox*{.95\textwidth}{!}{
\begin{tabular}{l | c | c | c | c | c | c} 
\hline\hline
 Pri. poly of $GF(2^8)$ & N.L & BIC & BIC of SAC & SAC & LP & DP \\ 
\hline

$x^8+x^4+x^3+x^2+1$ & 104.75 & 105.071 & 0.500 & 0.493 & 160/ 0.125 & 0.125 \\

$x^8+x^5+x^3+x^1+1$ & 105.75 & 104.929 & 0.502 & 0.503 & 158/ 0.140 & 0.242\\

$x^8+x^5+x^3+x^2+1$ & 104.75 & 101.14 & 0.502 & 0.497 & 168/ 0.156 & 0.5 \\

$x^8+x^6+x^4+x^3+x^2+x^1+1$ & 105.75 & 105.35 & 0.502 & 0.502 & 160/0.125 & 0.125  \\

$x^8+x^7+x^6+x^1+1$ & 104.5 & 104.14 & 0.498 & 0.498 & 164/0.148 & 0.25 \\

$x^8+x^6+x^5+x^2+1$ & 105.5 & 105.71 & 0.502 & 0.505 & 160/0.125 & 0.125 \\

$x^8+x^7+x^2+x^1+1$ & 106.75 & 104.85  & 0.503 & 0.502 & 160/0.125 & 0.125 \\

$x^8+x^7+x^3+x^2+1$ & 104.25 & 104.42 & 0.501 & 0.512 & 162/0.132 &  0.25 \\

$x^8+x^7+x^6+x^5+x^4+x^2+1$ & 106.5 & 105 & 0.504 & 0.496 & 162/0.132 & 0.117\\

$x^8+x^7+x^6+x^5+x^2+x^1+1$ & 106.25 & 103.71 & 0.500 & 0.498 & 162/0.132 & 0.125 \\

$x^8+x^6+x^3+x^2+1$ & 106 & 105.71 & 0.501  & 0.499 & 158/ 0.125 & 0.125 \\

$x^8+x^7+x^6+x^5+x^2+x^1+1$ & 106 & 103.57 & 0.502 & 0.497 & 166/0.156 & 0.25 \\

$x^8+x^6+x^5+x^1+1$ & 106.5 & 105.5 & 0.502 & 0.510 & 162/ 0.132 & 0.125 \\

$x^8+x^6+x^5+x^3+1$ & 106.25 & 105.37 & 0.504 & 0.507 & 158/0.132 & 0.125 \\

$x^8+x^6+x^5+x^4+1$ & 107.25 & 106.07 & 0.5 & 0.496 & 158/0.125 & 0.117 \\

$x^8+x^7+x^5+x^3+1$ & 106 & 105.35 & 0.503 & 0.516 & 162/0.132 & 0.125 \\
\hline

\hline
\end{tabular}
}
\caption{Analysis of proposed S-boxes with different Primitive polynomial of $GF(2^8)$.} 
\label{t_c} 
\end{table*}

\section{Proposed algorithm for image encryption cryptosystem}

For encryption of the digital material, a pseudorandom sequence generator needs to be developed for transforming the real outputs of chaotic Lorenz system into a digital sequence. This step is summarized below.
The Chaotic Lorenz system considered in this work can generate three real sequences which we denote as x, y, and z. For the purpose of enhanced randomness, we separate the first 100 values of the real sequences and denote the three real sequences as  
$\left\{x_i \right\},\left\{y_i\right\},\left\{z_i\right\}$, where $ i=1,2,3,...,m\times n$, where $m\times n$ is the size of plain image. The $t^{th}$ value of the sequences of $x$, $y$, and $z$ are defined as $x_t$, $y_t$ and $z_t$. We incorporate a disturbing procession in order to evade the appearance of the periodic absolutely. This is done by altering the value of $x$ and $y$ with an interval of $10000$:

\begin{eqnarray}\label{eqn4}
\begin{cases} 
      x_t \ = \ x_t \ + \ 0.1, \ y_t \ = \ y_t \ - \ 0.2,    & \quad if\ \ z_t \ \leq \ 0, \ t \ \equiv1 \ $\textbf{mod}$ \ (10000),\\
 			x_t \ = \ x_t \ + \ 0.2, \ y_t \ = \ y_t \ - \ 0.1,    & \quad if\ \ z_t \ > \ 0, \ t \ \equiv1 \ $\textbf{mod}$ \ (10000).\\
   \end{cases}
\end{eqnarray}
At first, the integral part of the real sequences for all of $x$, $y$ and $z$ is discarded:

\begin{eqnarray}
\begin{cases} 
      x_i=x_i-floor(x_i),\\
      y_i=y_i-floor(y_i), \     i \ = \ 1, \ 2, \ 3, \ ..., \ m \ \times \ n,  \\ 
			z_i=z_i-floor(z_i).\\
   \end{cases}
\end{eqnarray}
where $floor(x)$ means the maximum integer which is smaller than $x$.
Next, a new chaotic sequence is developed as follows:
\begin{eqnarray}
k_i={x_{i},y_{i},z_{i},x_{i+1},y_{i+1},z_{i+1},...,x_{i+floor(\frac{m\times n}{3})},y_{i+floor(\frac{m\times n}{3})}},z_{i+floor(\frac{m\times n}{3})}
\end{eqnarray}
At last, we get the modified chaotic sequence $k_i$. The beauty of proposed chaotic sequence is that it has the flavor of three $x_i$,$y_i$ and $z_i$ chaotic sequences of chaotic Lorenz system. During experiment we have changed the length of $k_i$ according to the size of plain image $m\times n$ by discarding some of its values from the end.

In order to get an improved image encryption scheme, we have changed the position of plaintext image pixels by randomness of $k_i$. The proposed scheme is a two phases scheme. 
The first phase starts from equation \ref{eqn4}  and ends at equation \ref{eqn8}. The basic purpose of first phase is to change the pixel positions of image. 
The second phase consists of equation \ref{eqn9}, \ref{eqn10} and \ref{eqn11} and in this process we are attaining  pixel value change based on XOR operation to get a fully encrypted image. 
Let us assume that the plain image is $I$ with $m$ rows and $n$ columns. For the sake of convinience, we have changed the image $I$ to a one dimensional vector, say $I_1(1:$m$\times$n$)$.
\begin{eqnarray}
 I_1((p-1)n+q)=I(p,q)
\end{eqnarray}
where $p=1,2,3,...,m$ and $q=1,2,3,...,n$. Now, the chaotic sequence $k_i$, will change the vector $I_1$ to $I_2$ with the help of following procedure.
\begin{eqnarray}\label{eqn8}
I_2(i)=I_1(k_i)
\end{eqnarray}
where $i=1,2,3,..., m\times n$. Define a new sequence with elements from $GF(2^8)$ based on $k_i$ as follows.
\begin{eqnarray}\label{eqn9}
l(i)= mod(round((k_i\times 10^{4})), 256)
\end{eqnarray}
After getting $l(i)$ sequence of random decimal numbers from $GF(2^8)$, we will $XOR$ $l(i)$ with $I_2(i)$ to get the final vector $I_3(i)$. Reshape $I_5(i)$ in the form of $m\times n$ matrix to get the first level ciphered image as shown in equation \ref{eqn11}.  
\begin{eqnarray}\label{eqn10}
I_3(i)= I_2(i)\oplus l(i)
\end{eqnarray}
\begin{eqnarray}\label{eqn11}
C.I= reshape(I_3(i), m, n).
\end{eqnarray}

As a last step for the proposed algorithm, we have applied the substitution step. We can define the process of substitution in following steps:

\begin{enumerate}
	\item	Consider the pixelof the image in the form of binary byte i.e., 8 binary bits. We divide this set into four LSBs (Least significant bits) and Most significant bits (MSBs).
	\item In the next step, the MSBs and LSBs having 4 bits each are converted into decimal values. Conventionally, the pixel value which has to be replaced by S-box value is selected with the help of decimal value of LSBs and MSBs.  The S-box column is selected by the decimal value of LSBs whereas the row of S-box is carefully chosen by decimal value of MSBs.
	\item One by one all the pixel values are substituted with the values of S-box.
	\item In this case, as there are number of S-boxes so each pixel of the image is replaced by single S-box, second pixel value is replaced by another S-box and this procedure will continue till last pixel value.
	\item For the selection of S-box out of 256 S-boxes with whom the original value will be replaced is done with the help of $x$, $y$, $z$ trajectories of Lorenz map.
\end{enumerate}

\section{Simulation results and statistical analysis}
For simulation results, we take an image of cameraman having size $256 \times 256 \times 3$. Table 4 represents the initial values of the chaotic maps which are used as secret keys. The plain image and histogram of cameraman are depicted in Fig. 1a and Fig. 1c. Fig.1a undergoes the process of encryption through our proposed encryption scheme. Fig. 1b gives the encrypted image. The quality of encryption algorithm is exhibited in the  strong visual results of encrypted image. 
Furthermore, the histogram of encrypted image in Fig.1d also confirms the strength of our scheme. The proposed image encryption scheme is also applied on gray scale image having gray value of 124. The gray scale image is given in Fig. 2a. 
 Fig. 2b represents the encrypted image of the gray scale image.  Moreover, Fig. 2c and Fig. 2d depict respectively the histogram images of plaintext image and encrypted image. The encryption results are satisfactory regardless that the plain image is high in autocorrelation. Furthermore, the histogram of the encrypted gray scale image show strength of our proposed technique. 

\begin{table}[h!]
\centering 
  \begin{tabular}{p{3cm} p{1cm} p{1cm} p{1cm}}
     
			\hline
				Parameters			& \(a\) & \(b\) & \(c\) \\
			\hline \hline
												& 10 & 28 & $\frac{8}{3}$\\
			\hline
  \end{tabular}
\caption{The initial conditions for the chaotic maps utilized as the secret keys in proposed encryption technique.} 
	\label{t_2}
\end{table}

\begin{figure*}

\centering
\subfloat[]{\includegraphics[width=40mm]{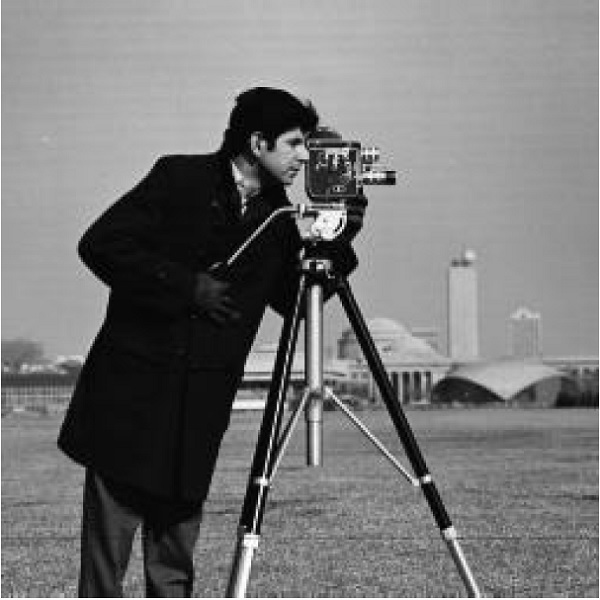} }
\centering
\subfloat[]{\includegraphics[width=40mm]{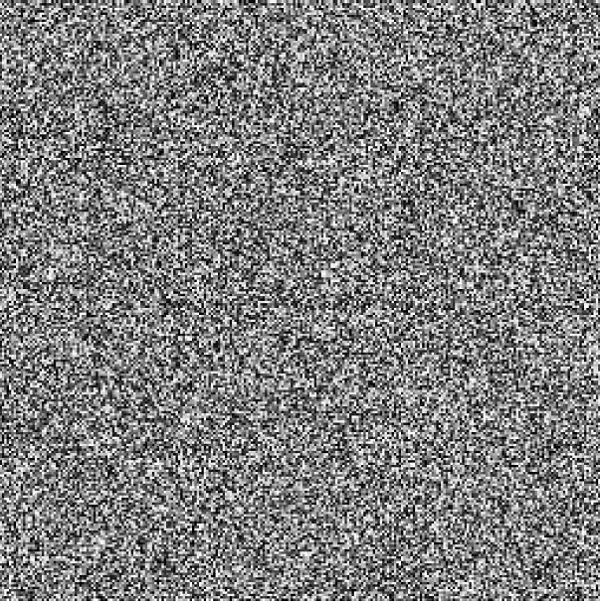} }
\centering
\subfloat[]{\includegraphics[width=47mm]{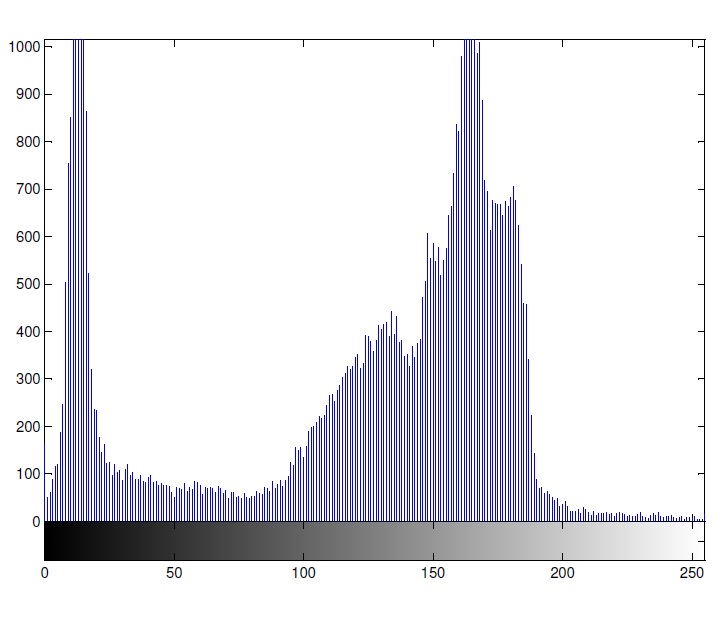} }
\centering
\subfloat[]{\includegraphics[width=47mm]{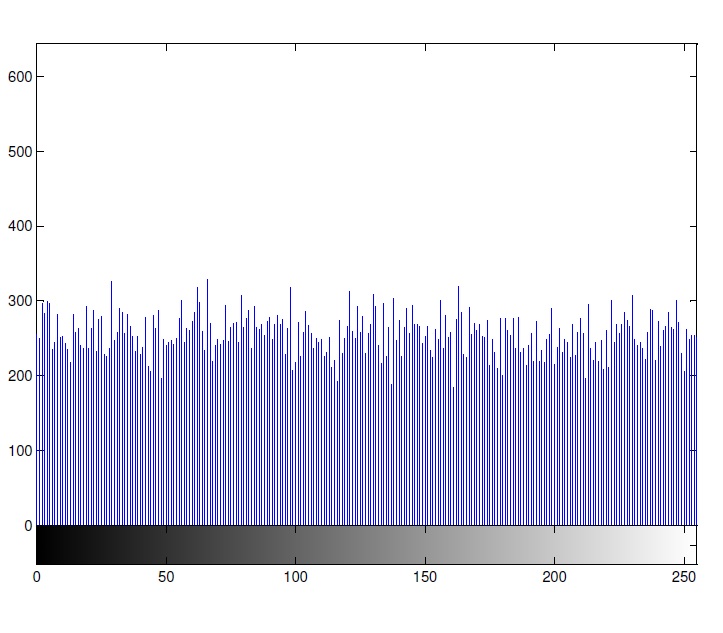} }

\caption{Simulation outcomes of proposed scheme (a) $256 \times 256$ size plain image of cameraman (b) cameraman image after encryption with secret keys (c) histogram analysis of cameraman (d) histogram analysis of encrypted cameraman.}
\label{lena}
\end{figure*}

\begin{figure}
  \centering
	  \subfloat[]{\includegraphics[width=40mm]{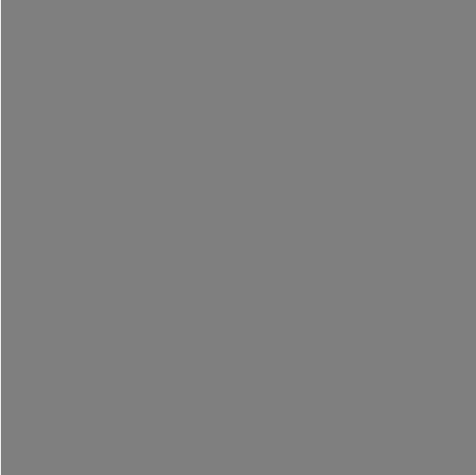} } 
		\subfloat[]{\includegraphics[width=40mm]{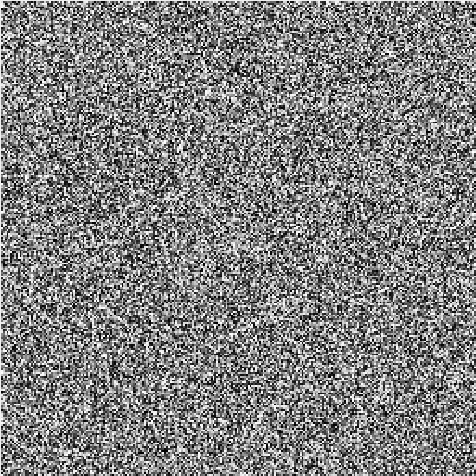} }
		\subfloat[]{\includegraphics[width=47mm]{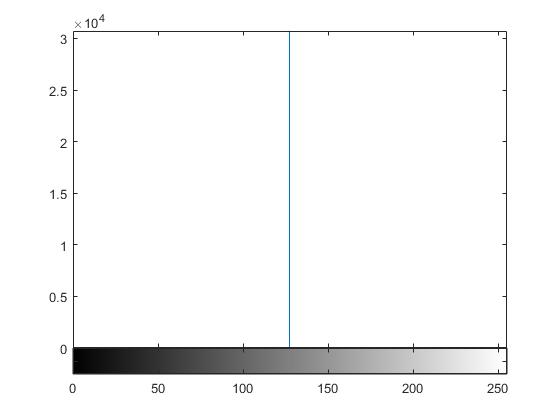} }  
		\subfloat[]{\includegraphics[width=47mm]{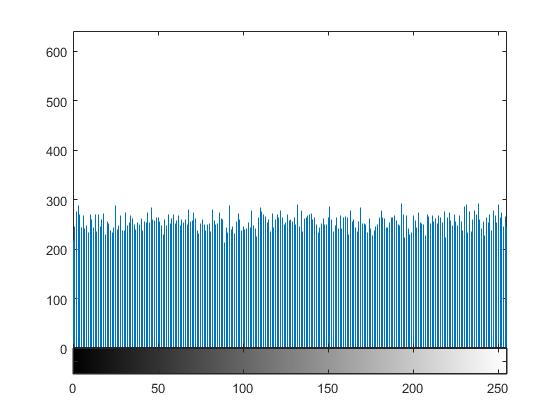} }
	  \caption{Simulation outcomes of new image encryption technique. (a)  gray scale image having  $256 \times 256$ size, (b) encrypted image (c) histogram results of one gray scale image (d) histogram results of encrypted one gray scale image.}
	\label{one_encr}
\end{figure}

\subsection{Statistical analysis}
The purpose of  statistical analysis is for assessing the robustness of our designed image encryption technique. 
We employ distinct statistical analyses to determine the standard of our proposed image encryption technique. In addition to this, the outcomes of our proposed technique are compared with the corresponding results of well-known image encryption methods. The description of these analyses is given in the following subsections.

\subsubsection{Correlation}

Given an image with  \(\rho(i, j)\) representing the pixel value at $(i,j)$ position of the image.  The correlation of  image is defined by \cite{r30} :

\begin{equation}
Corr. = \sum_{i, j} \frac{(i - \mu i)(j - \mu j)\rho(i, j)}{\varphi_{i}\varphi_{j}}. 
\end{equation}

where the standard deviation is denoted by $\varphi$ and $\mu$ is the variance.  The value of correlation provides a measure of  similarity between the two neighboring pixels over the entire image. Its values span is in $[-1 \ 1]$ where 1 represents  perfect correlation.

\begin{figure*}
\centering
\subfloat[]{\includegraphics[width=80mm]{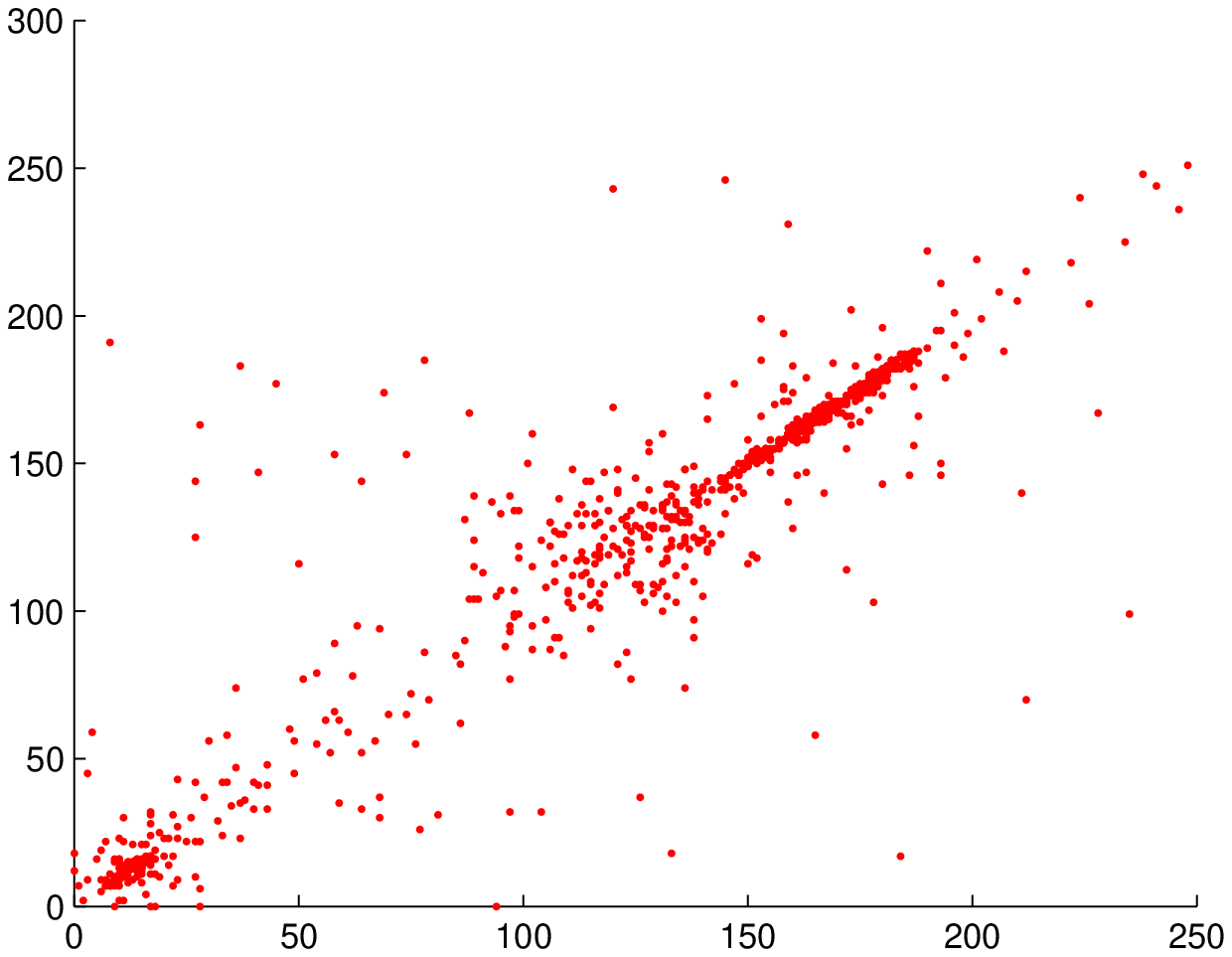} }
\centering
\subfloat[]{\includegraphics[width=80mm]{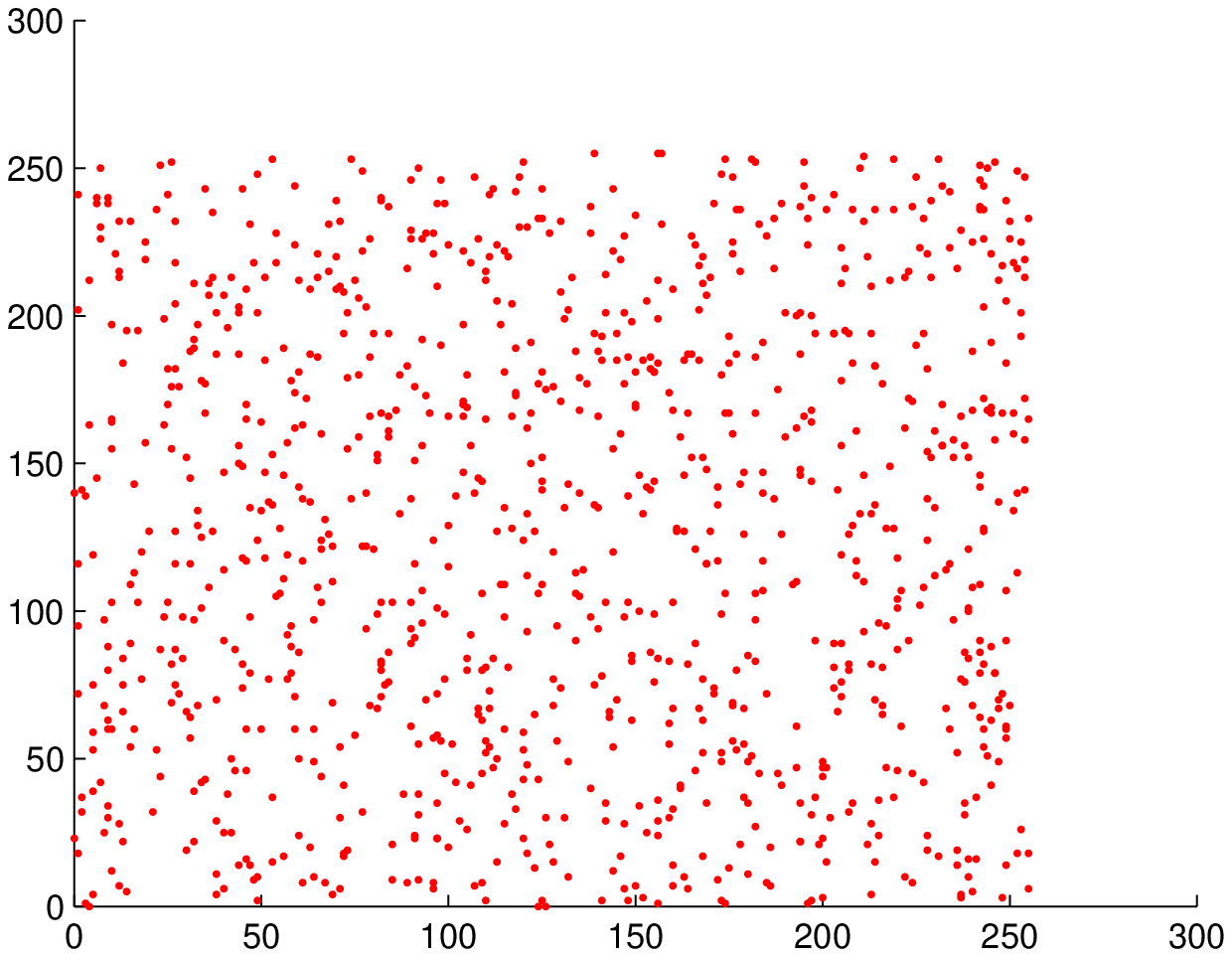} } 
\caption{ The distribution of horizontally adjacent pixels of (a) cameraman image and (b) encrypted cameraman image. The distribution of vertically adjacent pixels of ciphered cameraman image will have same behavior as pixels that are horizontal adjacent .}
\label{f_h_dis}
\end{figure*}

The distribution of horizontally adjacent pixels of cameraman image is shown in Fig.\hspace{1mm}\ref{f_h_dis}a.  
Fig.\hspace{1mm}\ref{f_h_dis}b displays the distribution for those pixels of encrypted cameraman image which are horizontally adjacent. The distribution of vertical adjacent pixels in ciphered cameraman image will act similar as they respond in horizontal adjacent pixels.

\subsubsection{Entropy}

Given an image with  \(\rho(i, j)\) representing the pixel value at $(i,j)$ position of the image.  The entropy of image is defined by  \cite{r30}:

\begin{equation}
Entropy = -\sum_{i, j} pr(\rho(i, j)) \log_{2} pr(\rho(i, j)). 
\end{equation}

where  \(pr(\rho(i, j))\) denotes the probability of image pixel.  For an image having 256 gray scales, the entropy values are distributed in \([0 \ 8]\)  and determine the randomness of image. Bigger values of entropy imply bigger measure of randomness. 

\subsubsection{Contrast}

Given an image with  \(\rho(i, j)\) representing the pixel value at $(i,j)$ position of the image.  The contrast of  image is defined by  \cite{r30}:

\begin{equation}
Contrast = \sum_{i, j} |i - j|^{2} \rho(i, j). 
\end{equation}

The contrast values have the range \([0 \ (size(Image)-1)^2]\). In case of a constant image, the contrast value is 0. Bigger values of contrast imply more variation in image pixels. The image contrast analysis  makes it possible for the observer to clearly recognize the objects in the image texture.   

\subsubsection{Homogeneity}

In  \cite{r30}, the homogeneity of image is defined by:

\begin{equation}
Homo. = \sum_{i, j}\frac{\rho(i, j)}{1 + |i - j|}.
\end{equation}

where the location of image pixels is given by  \(i, j\) . In this analysis, the closeness of gray level cooccurrence matric (GLCM) diagonal and GLCM is calculated. The homogeneity values lie in the interval  \([0 \ 1]\).

\subsubsection{Energy}
The energy of an image can be defined by
\begin{equation}
Energy = \sum_{i, j} \rho(i, j)^{2}. 
\end{equation}
where \(i, j\) depicts the position of image pixels. The above equation interprets the energy as the summation of the square of all elements of GLCM. The  energy values lies in the interval \([0 \ 1]\) and the constant image has maximum energy value of 1. 

Table\hspace{1mm}\ref{t_3} shows the values of statistical analyses and also the comparison with other existing techniques. This comparison indicates the quality of our proposed scheme.

\begin{table}[h!]
\centering 
\resizebox*{.95\textwidth}{!}{
\begin{tabular}{p{3cm} | p{2cm} p{2cm} p{2cm} p{2cm} p{2cm}} 
\hline\hline
Analysis & Corr. & Entropy & Homo. & Contrast & Energy  \\ 
\hline\hline 
Ref. \cite{r34} & -0.0308 & 7.9311 & 0.8365 & 8.0522 & 0.1984       \\
Ref. \cite{r32} & 0.0439 & 2.5643 & 0.5733 & 4.9454 & 0.4263       \\
Ref. \cite{r35} & -0.0293 & 7.9801 & 0.9102 & 8.6603 & 0.0674       \\
Ref. \cite{r33} & 0.0313 & 7.9735 & 0.8251 & 8.1833 & 0.2132       \\
Ref. \cite{r31}  & 0.0687 & 7.1735 & 0.8121 & 8.3849 & 0.1254       \\
\textbf{Proposed} & \textbf{-3e-4} & \textbf{7.9521} & \textbf{0.9598} & \textbf{8.4587} & \textbf{0.3521}       \\
\hline 
\end{tabular}
}
\caption{Comparison of the statistical analyses of the encrypted images of Lena obtained through application of proposed image encryption algorithm and other relevant techniques.} 
\label{t_3} 
\end{table}

\section{Security analysis}

It is mandatory to compute the security analyses to assess the strength of any cryptosystem. Here, we assess the security of our scheme with the help of certain security analyses like key space, key sensitivity, avalanche analysis, noise resistant analysis and cryptanalysis. The comparison of the outcomes of these analyses with the security analyses of other schemes exemplify the strength of our proposed technique. Following sections describe the security analyses in detail.

\subsection{Key space and key sensitivity}
For any cryptographic system, the total count of secret keys which are utilized to encrypt the data are named as key space and it has its importance as far as security of whole scheme is concerned. In this work, the initial conditions of three different chaotic maps are  the secret keys. The secret key has average range of $10^{20}\) and we have used three different secret keys so the total count of different keys is given as \(10^{20\times 3} = 10^{60}\). A recent personal computer will require over $10^{10}$ years to go through all possible blends of this huge keyspace.

\begin{figure}
  \centering
	  \subfloat[]{\includegraphics[width=0.33\textwidth]{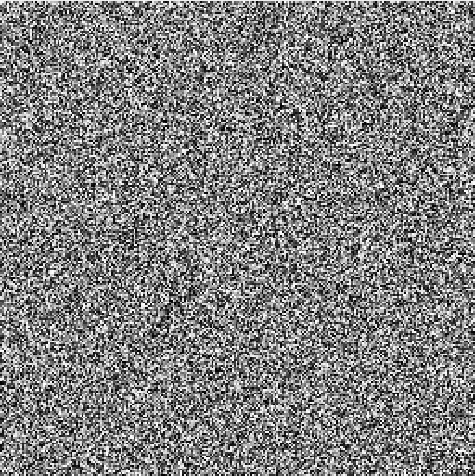} } 
		\subfloat[]{\includegraphics[width=0.33\textwidth]{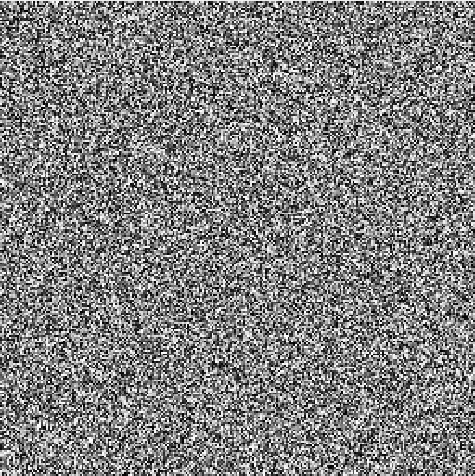} }\\
		\subfloat[]{\includegraphics[width=0.33\textwidth]{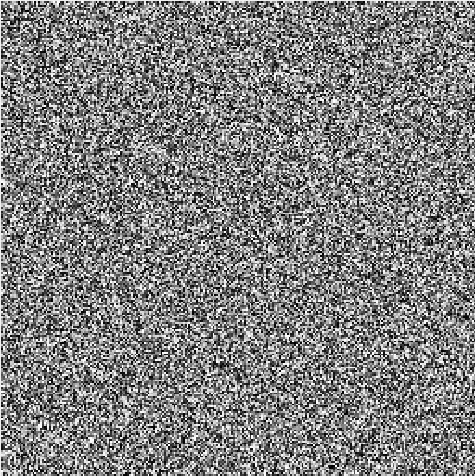} } 
		\subfloat[]{\includegraphics[width=0.33\textwidth]{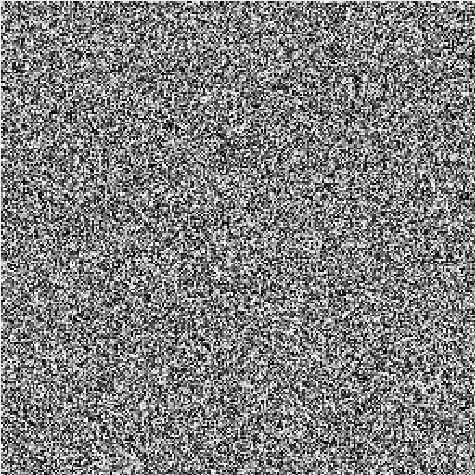} }
	  \caption{Key sensitivity analysis (a) First secret key is altered from  \(k_1 = a = 10\) to \(k_{1}^{'} = a = 10.0000000001\) (b)Second secret key is \(k_2 = b = 28\) to \(k_{2}^{'} = b = 28.0000000001\) (c) Third secret key is altered from   \(k_3 = c = \frac{8}{3}  \) to \(k_{3}^{'} = c = \frac{8}{3} + 0.0000000001\) (d) Fourth secret key is altered from  \(k_1 = a = 10\) to \(k_{1}^{'} = a = 9.99999999999\).}
	\label{key_sen}
\end{figure}

One of the features of the quality cryptosystem is its sensitivity to a tiny change in secret keys. For instance, the change in secret key during decoding process will give altogether a different decoded image. This mechanism is named as key sensitivity. Our designed encryption technique is sensitive even to a minor variation in the initial conditions. To prove this claim, we will change the secret keys and show the pictorial results. By using secret keys given in Table \hspace{1mm}\ref{t_2}, we have encrypted the cameraman image as given in Fig. \hspace{1mm}\ref{lena}b. Considering four different cases of changing initial secret keys we have following results.\\
\textbf{Case I}: If the initial secret key $k_1$ is slightly changed i.e., \(k_1 = a = 10\) to \(k_{1}^{'} = a = 10.0000000001\) then it is observed that the decryption process does not get the required results. Fig. 4a depicts the decryption of plain-image with key \(k_{1}^{'}\). In this process, the remaining two keys were remained the same.\\
\textbf{Case II}: For the second case, the key \(k_{2}\) is changed from its original value i.e., \(k_2 = b = 28\) to \(k_{2}^{'} = b = 28.0000000001\) and again with this slight change in one key, the decryption  image does not resemble to the original plaintext image and hence prove our claim of key sensitivity. The decryption is given in Fig. \hspace{1mm}\ref{key_sen}b. \\
\textbf{Case III}: The initial secret key \(k_3 \) is slightly changed while keeping the remaining keys same. A change of 0.0000000001 in \(k_{3}\) i.e \(k_3 = c = \frac{8}{3} \) to \(k_{3}^{'} = c = \frac{8}{3} + 0.0000000001\) provides a different original image as given in Fig. \hspace{1mm}\ref{key_sen}c. \\
\textbf{Case IV}: For the last case, the key \(k_4 \) is changed like \(k_1 = a = 10\) to \(k_{1}^{'} = a = 9.99999999999\) and the decrypted image is given in Fig. \hspace{1mm}\ref{key_sen}d. \\
For each of the four cases, even a slight variation in the initial secret key could not obtain the original image and hence proving our claim of key sensitivity for proposed algorithm.

\subsection{Avalanche analysis}

In block ciphers, the effect of avalanche mentions one of the properties of strong cryptographic algorithms. If a single input bit change effect the half number of output bits, then it is an apparent avalanche effect. Researchers prefer unified average change intensity (UACI) and number of pixel change rate (NPCR) to measure the effect of avalanche criteria. The detailed description of this effect is provide in  \cite{r36}  as 
\begin{equation}
	NPCR = \frac{\sum_{i,j}D(i,j)}{N \times M}\times100 \%,
\end{equation}
\begin{equation}
	UACI = \frac{1}{N \times M}\left[\sum_{i,j}\frac{|C_1(i,j) - C_2(i,j)|}{255}\right] \times 100 \%,
\end{equation}
where the two ciphered digital images \(C_1\) and \(C_2\) are attained by changing the single bit of plain image. Moreover, the height and width of cipher images are given by \(N\) and \(M\) respectively. We can define the \(D(i,j)\) as
	\[ D(i,j) = \left\{ 
  \begin{array}{l l}
    0 & \quad \text{if \(C_1(i,j) = C_2(i,j)\)},\\
    1 & \quad \text{if \(C_1(i,j) \neq C_2(i,j)\)}.
  \end{array} \right.\]

	By adjusting the single pixel of plaintext image, the rate of change of pixel quantity of encrypted image is measured by number of pixel change rate (NPCR) analysis. Moreover, the normal power of contrast between plain and encrypted images is calculated by unified average change intensity analysis (UACI). The calculated minimum value for NPCR must be $50 percent$. In our work, three different plain images of Lena, baboon and cameraman have been through NPCR and UACI analyses. For each image, the position of bit is firstly changed around first pixel then around middle pixel and finally around the last pixel. Table 6 depicts the outcomes of NPCR and UACI for all cases of three images. In this Table the value of NPCR remains greater than 99 percent and value of UACI is greater than 33 percent. These results indicate the strong avalanche effects. In addition to this, a comparison has been established between avalanche values of proposed scheme and AES
	
	\begin{table}[h]
\centering 
  \begin{tabular}{p{.9cm} p{1.1cm} p{1cm} p{1cm} p{1cm} p{1cm}}
    \hline \hline
      \multicolumn{2}{c}{Analysis} &
      \multicolumn{2}{c}{UACI(\%)} &
      \multicolumn{2}{c}{NPCR(\%)} \\
			\hline
			\multicolumn{2}{c}{Images \& Loc.} 	& Prop. 		& AES 			& Prop. 		& AES \\
			\hline \hline
		\multirow{3}{*}{Baboon} 	& first 			& 33.3620 	& 33.4463  & 99.1252 	& 99.6124  \\
															& mid 			 	& 33.4510 	& 33.4561     & 99.5271 	& 99.6033 \\
															& last 			 	& 33.6930 	& 33.5252  & 99.6389 	& 99.6185 \\
  \hline
		\multirow{3}{*}{Lena} 		& first 		 	& 33.2010 	& 33.3996    & 99.2563 	& 99.6094  \\
															& mid 			 	& 33.6320 	& 33.3139  & 99.5210 	& 99.6506 \\
															& last 				& 33.5202 	& 33.5133  & 99.4198 	& 99.6002  \\
    \hline
 \multirow{3}{*}{Cman} 		& first 		 	& 33.5120 	& 33.5360  & 99.5212 	& 99.6048  \\
															& mid 				& 33.3014 	& 33.5212  & 99.5862 	& 99.6201   \\
															& last 			 	& 33.4120 	& 33.5245  & 99.5410 	& 99.5819 \\
    \hline
		\multirow{2}{*}{Key S.} 	& Case I 			& 33.7401 	& 33.5029 & 99.4802 	& 99.5972  \\
															& Case II 	 	& 33.2015 	& 33.5468  & 99.0025 	& 99.6460 \\
    \hline
  \end{tabular}
\caption{The comparison of UACI and NPCR analysis of our designed technique and AES on images of Lena, Baboon and cameraman.  Three cases are proposed for each image which are changing of single bit in the first pixel, the mid pixel and the last pixel. Moreover, the key sensitivity analysis for two cases is given.} 
	\label{t_npcr}
\end{table}

By using NPCR and UACI analyses, the key sensitivity of proposed scheme is also evaluated. In first case, two same keys with difference of only 1 bit are used to calculate the difference between two encrypted images. For the second case, the difference of one bit between two keys would remain the same but we calculated the difference between encrypted and decrypted images. Table \hspace{1mm}\ref{t_npcr} gives the results of both cases along with comparison with AES. The results of Table show the required avalanche effect.

\subsection{Noise resistant analysis} 

The noise resistant encryption algorithm depicts the strength of any cryptosystem. Any transmitted data may get effected by channel (irrespective of wired or wireless channel) noise or deliberately added noise. It is observed that it is hard to decipher the abandoned cipher image even the portion of the image is affected. Few methods like error detection and correction have been employed to counter these situations but at the cost of computational complexity. For successful transmission and decryption of cipher data, the error detection and correction are required before both the steps. So, ultimately it increases the complexity of the system.
In this proposed technique, the addition of noise does not become the hurdle to decipher image correctly with some minor changes. To verify this claim, a series of experiments regarding successful deciphering have been done with noise addition in the cipher images. Fig. 5a and Fig 5b. represent the plain image and encrypted images respectively. This encryption is done with the help of secret keys of Table 4. For noise resistance test, the 10,000 pixels of encrypted image are either cropped or made corrupted with white pixels as illustrated in Fig. 5c. The deciphering of this image is given in Fig. 5d. Clearly, this pictorial representation indicates the successful decryption with minor changes. To prove this claim for other images, the cameraman image is encrypted with the secret keys of Table 4. The plain image with noise and its encryption is given in Fig. 5e and Fig. 5f respectively. Now, again we removed first 10, 000 pixels either with the help of cropping or by corrupting the image. Interestingly, the decryption is successful with minor changes and is shown in Fig. 5h.
On the other hand, there is a difference between the deciphering of plain text and digital image. For noisy cipher text, the deciphering of the text gives a whole new text. There are many examples in which the major concern is to recognize the face of the person and object irrespective of quality of decipher images. For this reason, the deciphering of encrypted images having added noise is major breakthrough.
\begin{figure*}
\centering
\subfloat[]{\includegraphics[width=160mm]{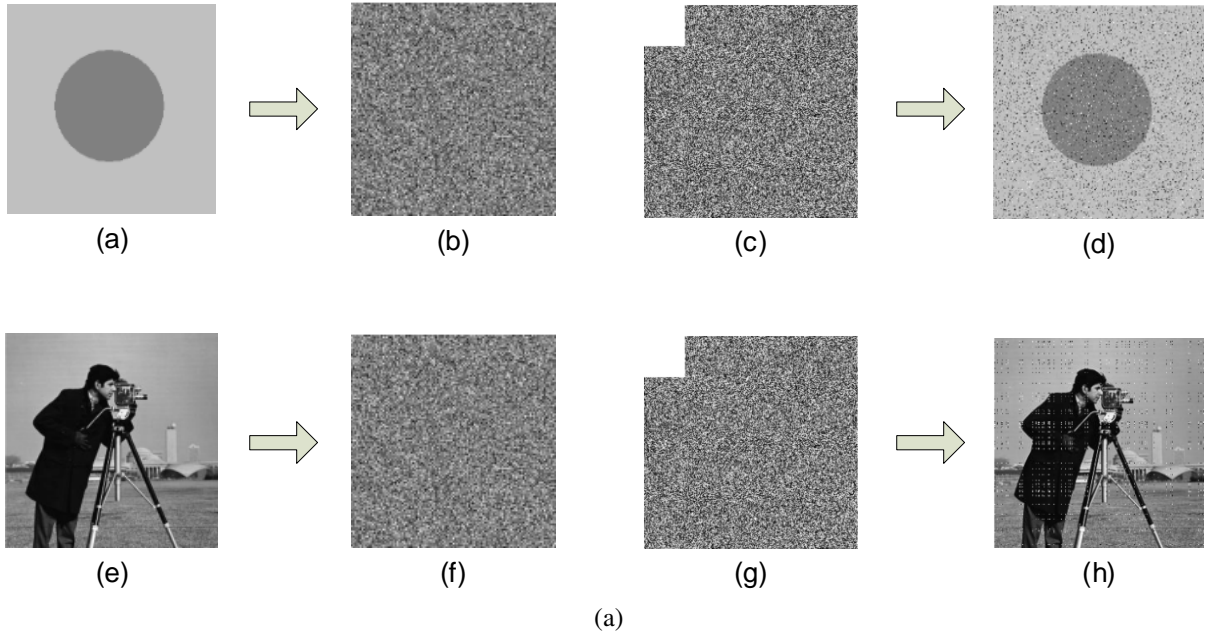} }
\caption{(a) A test image, (b) encryption of test image using secret keys. (c) Adding noise in the cipher image by changing first 10,000 pixels either by cropping or corrupting the file by using the white pixels. (d) Deciphered image. (e) The cameraman noisy image, (f) encryption of cameraman image using secret keys. (g) Adding noise in the cipher image by changing first 10,000 pixels either by cropping or corrupting the file by using the white pixels. After decryption, (h) Deciphered image.}
\label{noise}
\end{figure*}
\subsection{Cryptanalysis}
To assess the strength of proposed scheme against different malicious attacks, following attacks are considered to evaluate the proposed cryptosystem.
\subsubsection{Linear cryptanalysis}
Linear approximation probability is used to analyze the imbalance of an event. The maximal amount of imbalance of the event can also be obtained with the help of this analysis. In this analysis, two masks $\Gamma x\) and $\Gamma y\), respectively are applied to parity of both input and output bits. In  \cite{r37} , it is defined as
\begin{equation}
	LP = max_{\Gamma x \Gamma y \neq 0}\left|\frac{\left\{x/X \bullet \Gamma x = S(x)\bullet \Gamma y = \Delta y\right\}}{2^{n}} - \frac{1}{2}\right|,
\end{equation}
where all the inputs values are contained by set \(X\) and total number of this set are  \(2^{n}\). As we have used 8 different S-boxes and the average maximum value of LP is \(LP_{max} = 2^{-4.21}\) and having distinct 256 S-boxes the maximum LP is given as \(LP_{max}^{4r} = 2^{-4.21 \times 256} = 2^{-1077}\). By seeing the outcomes of LP, it is almost impossible for an invader to differentiate our proposed cipher with the help of random permutation and hence, the proposed algorithm will show resistance for countering linear cryptanalysis. 
\subsubsection{Differential cryptanalysis}
One of the precise goals for assurance of uniform mapping, the input differential extraordinarily supervises the differential at output end. These features certify the probability of uniform mapping for every input bit $i$. The main purpose of approximation probability is to calculate differential uniformity of S-box. In  \cite{r38}   it is given as:
\begin{equation}
	DP(\Delta x\rightarrow \Delta y) = \left[\frac{\left\{x\in X /S(x)\oplus S(x\oplus \Delta x) = \Delta y\right\}}{2^{m}}\right],
\end{equation}
where the input and output differentials are given by $\Delta x\) and $\Delta y\) respectively. For S-boxes, the maximum average value of  DP is   \(DP_{max} = 2^{-4.05}\). Here, the number of active S-boxes is exactly 256 i.e.,  \(DP_{max}^{4r} = 2^{-4.05 \times 256} = 2^{-1036}\). By seeing this outcome, it is confirmed  that proposed scheme has the ability to resist against malicious differential cryptanalysis.
\section{Conclusion}
In this paper, a combination of proposed S-boxes and chaotic maps is used for image encryption algorithm. This scheme consists of two phases. In the first phase, substitution is performed via multiple S-boxes instead of a single S-box. The application of several S-boxes not only provides additional security but also utilizes less rounds of encryption. The initial values of Lorenz chaotic map help to operate each round of encryption. In addition to this, permutation is performed in the second phase. The resistance of proposed encryption scheme is depicted through simulation and security analyses results. The outcomes of cryptanalysis also confirm the robustness of our proposed scheme. This work also motivates researchers to add different changes like increasing number of S-boxes and encryption rounds while keeping standard of encryption and computational complexity.

\end{document}